\documentclass[twocolumn]{aastex7}

\usepackage{float}
\usepackage{natbib}
\usepackage{hyperref}
\usepackage[nointegrals]{wasysym}
\usepackage{ragged2e}
\usepackage{graphicx}
\usepackage{subcaption}
\usepackage{units}
\usepackage{amsmath}
\definecolor{red}{rgb}{1,0,0}
\definecolor{orange}{RGB}{204, 85, 0}
\definecolor{blue}{HTML}{4169e1}
\definecolor{ltred}{RGB}{245,167,162}
\definecolor{ltblue}{RGB}{206,211,242}

\newcommand{\ha}{H$\alpha$}

\newcommand{\hb}{H$\beta$}

\newcommand{\oiii}{[O\,{\sc iii}]}

\newcommand{\Msun}{M$_\odot$}

\newcommand{\hst}{\textit{HST}}
\newcommand{\jwst}{\textit{JWST}}

\defcitealias{roberts-borsani_borg-jwst_2025}{RB25}
\defcitealias{rojas-ruiz_probing_2020}{RR20}
\defcitealias{leethochawalit_uv_2023}{L23}
\defcitealias{rojas-ruiz_borg-jwst_2025}{RR25}

\shorttitle{SALSA: a Faint and Strongly Ionizing source at $z=5.66$}
\shortauthors{Rojas-Ruiz et al.}

\begin{document}

\title{Chasing Cosmic Reionization: An Extremely Faint Highly Magnified Source at $z=5.66$ with high $\xi_{\rm ion}$.}

\correspondingauthor{Sof\'ia Rojas-Ruiz}
\email{rojas@astro.ucla.edu}

\author[0000-0003-2349-9310]{Sof\'ia Rojas-Ruiz}\affiliation{Department of Physics and Astronomy, University of California, Los Angeles, 430 Portola Plaza, Los Angeles, CA 90095, USA}
\email{rojas@astro.ucla.edu}

\author[0000-0003-1427-2456]{Matteo Messa}\affiliation{INAF - OAS, Osservatorio di Astrofisica e Scienza dello Spazio di Bologna, via Gobetti 93/3, I-40129 Bologna, Italy}
\email{matteo.messa@inaf.it}

\author[0000-0002-5057-135X]{Eros Vanzella}\affiliation{INAF - OAS, Osservatorio di Astrofisica e Scienza dello Spazio di Bologna, via Gobetti 93/3, I-40129 Bologna, Italy}
\email{eros.vanzella@inaf.it}

\author[0000-0002-8460-0390]{Tommaso Treu}
\affiliation{Department of Physics and Astronomy, University of California, Los Angeles, 430 Portola Plaza, Los Angeles, CA 90095, USA} 
\email{tt@astro.ucla.edu}

\author[0000-0003-1383-9414]{Pietro Bergamini}\affiliation{INAF - OAS, Osservatorio di Astrofisica e Scienza dello Spazio di Bologna, via Gobetti 93/3, I-40129 Bologna, Italy}
\email{pietro.bergamini@inaf.it}

\author[0009-0001-9416-0923]{Giorgia Di Rosa}\affiliation{Dipartimento di Fisica e Scienze della Terra, Università degli Studi di Ferrara, Via Giuseppe Saragat 1, 44122 Ferrara, Italy}\affiliation{INAF - OAS, Osservatorio di Astrofisica e Scienza dello Spazio di Bologna, via Gobetti 93/3, I-40129 Bologna, Italy}\email{giorgia.dirosa@unife.it}

\author[0000-0002-3336-4965]{Marco Lombardi}\affiliation{Dipartimento di Fisica "Aldo Pontremoli", Università degli Studi di Milano, Via Celoria 16, 20133 Milano, Italy}\affiliation{INAF - OAS, Osservatorio di Astrofisica e Scienza dello Spazio di Bologna, via Gobetti 93/3, I-40129 Bologna, Italy}
\email{marco.lombardi@unimi.it}

\author[0000-0002-6813-0632]{Piero Rosati}\affiliation{Dipartimento di Fisica e Scienze della Terra, Università degli Studi di Ferrara, Via Giuseppe Saragat 1, 44122 Ferrara, Italy}\affiliation{INAF - OAS, Osservatorio di Astrofisica e Scienza dello Spazio di Bologna, via Gobetti 93/3, I-40129 Bologna, Italy}\email{rstpri1@unife.it}

\begin{abstract}
We present \jwst\ NIRSpec IFU spectroscopic measurements of one of the faintest (M$_{UV}>-13.6$) known ionizing sources at $z=5.66$, dubbed Small And Lensed Source Arc, SALSA. This source is highly magnified ($\mu>100$) by the lensing galaxy cluster Abell 2744, providing a unique opportunity to investigate the physical properties of faint sources in the Epoch of Reionization. We characterize SALSA's nebular emission using rest-frame UV and optical emission lines and investigate the relationship between its ionizing efficiency, nebular excitation, and chemical enrichment. We robustly detect H$\alpha$ emission and use it to predict the H$\beta$ flux assuming negligible dust attenuation. We also measure the\oiii\,$\lambda$5007 emission and estimate oxygen abundance resulting in a high R3 index $2.82_{-0.25}^{+0.34}$ and relatively low metallicity $\mathrm{12+log(O/H)}= 7.43\pm0.09$. SALSA presents a high ionizing production efficiency log$(\xi_{ion})=25.49_{-0.08}^{+0.09}$ Hz erg~$^{-1}$, consistent with theoretical models from very massive stars and constant star formation rate, and a high Ly$\alpha$ escape fraction $f_{\rm{esc}}^{\mathrm{Ly}\alpha}=0.39\pm0.14$. These properties place SALSA among the most extreme star-forming sources known at this epoch.
\end{abstract}


\section{Introduction} 
Studying the physical conditions of the first galaxies is essential to understanding early galaxy assembly and evolution. These primitive galaxies are characterized as young, compact systems with extreme UV-ionizing power from metal-poor stars \citep[e.g.,][]{rosdahl_lyc_2022}, and responsible for Cosmic Reionization. This epoch represents the last major phase transition in the universe when the intergalactic medium (IGM) becomes transparent to Lyman continuum (LyC) radiation (energy $\geq13.6$ eV). Observations suggest that the bulk of reionization is done by $z\sim6$ \citep[e.g.,][]{loeb_reionization_2001, fan_constraining_2006,mcquinn_evolution_2016,yang_measurements_2020}. However, recent studies on quasar spectra show that this transition is not fully completed until $z\sim5.3$ \citep[e.g.,][]{bosman_hydrogen_2022,davies_updated_2026}. Although it is thought that galaxies contribute more than the more powerful but rare quasars and AGN \citep[e.g.,][]{robertson_discovery_2022} to cosmic reionization, a complete picture is still missing. For example, it is still unclear whether the relative contribution to reionization is from the brighter and more massive galaxies \citep[e.g.,][]{naidu_rapid_2020}, or from the more common faint and low-mass systems \citep[e.g.,][]{atek_extreme_2018,finkelstein_conditions_2019,atek_most_2024,simmonds_low-mass_2024}.

Recent \jwst\ studies have revealed a population of faint high-redshift galaxies with strong nebular emission and elevated ionizing photon production efficiencies, $\xi_{\rm ion}$, which could be a major contributor to Cosmic Reionization. It is thus important to investigate the relationship between the ionizing radiation field, nebular excitation, and chemical enrichment of this faint population \citep[e.g.,][]{vanzella_extremely_2023,vanzella_extreme_2024,vanzella_pristine_2026,atek_glimpse_2026,nakajima_ultra-faint_2026}. Strong gravitational lensing magnification, coupled with the sensitivity and resolution of \jwst, enables spectroscopy of these faint sources. This in turn provides a unique opportunity to push the limits of our understanding on the contribution of faint sources towards Reionization and early galaxy assembly.

In this work, we present \jwst/NIRCam and NIRSpec observations, as well as archival data from the Very Large Telescope Multi-Unit Spectroscopic Explorer (VLT/MUSE) of an exceptionally faint (M$_{\rm UV}>-13.6$ and highly ionizing source (dubbed Small And Lensed Source Arc, hereafter SALSA) found behind the Hubble Frontier Fields cluster Abell 2744 \citep{lotz_frontier_2017} at $z=5.66$.  We estimate $\xi_{\rm ion}$ and find it to be at the upper end of what is commonly assumed for sources at these redshifts. Furthermore, we use the R3 diagnostic ratio (\oiii\,$\lambda$5007/H$\beta$) as a probe of nebular excitation and gas-phase metallicity. We find it to be higher than expected at this UV luminosity \citep{korber_glimpse_2026}. We also use R3 to estimate Oxygen abundance, finding it to be approximately 5\% solar. The low metallicity is consistent with the elevated R3, given that high electron temperatures enhance \oiii\ emission \citep{nakajima_empress_2022}. 

This Letter is organized as follows: Section \S\ref{data} presents the spectroscopic and photometric data from MUSE and \jwst\ used in this work as well as details on the NIRSpec IFU data reduction. Section \S \ref{lens} describes the lens model we adopt for the analysis, and the physical calculations of the system are presented in \S \ref{analysis}. Finally the results and their interpretation in context with the literature are described in \S \ref{ionizing} and summarized in \S\ref{summary}. Throughout this work we use the cosmology according to H$_0$= 70 km~s$^{-1}$ Mpc$^{-1}$, $\Omega_M$= 0.3, and $\Omega_\Lambda$ = 0.7. All magnitudes are given in the AB system.

\section{Data}\label{data}
We use a multi-wavelength spectroscopic and photometric dataset to characterize the ionizing power of the SALSA arclet. For spectroscopy, we primarily rely on NIRSpec IFU PRISM observations from \jwst\ cycle 4 program GO 7677 (Co-PIs: E. Vanzella, M. Messa) targeting both the $z=4.19$ Lensed And Pristine-2 (LAP2) doubly imaged source in the Abell 2744 Cluster (\citealp{vanzella_pristine_2026}; Messa et al.~2026, in prep.), and the arclet SALSA studied here at $z=5.66$ with coordinates R.A. (J2000) $= 0^\mathrm{h} 14^\mathrm{m} 21\fs8678$, DEC. (J2000) $= -30^{o}23'56\farcs241$. The observations aim to investigate the ionizing power of these sources by tracing their ionized gas through emission lines such as \oiii\ and H$\alpha$. The total exposure time on target is 13.62 hours and the read out mode is NRSIRS2. More details on the observations are highlighted in Messa et. al (2026, in prep.) 

\begin{figure*}[ht!]
\centering
\includegraphics[width=\linewidth]{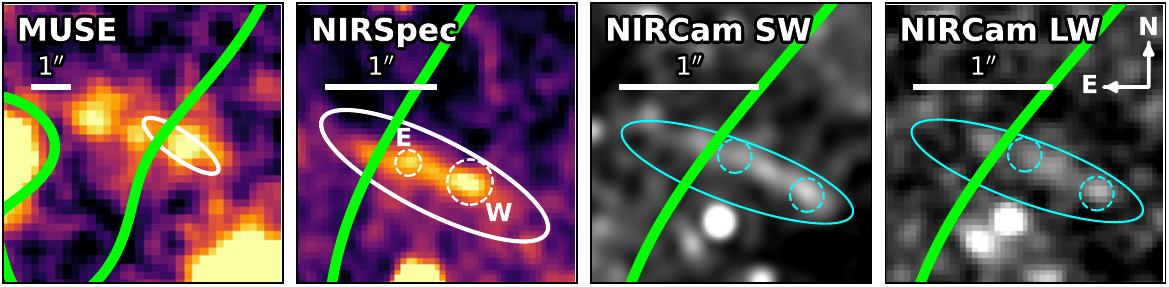}
\caption{Spatial morphology of SALSA and its position relative to the lensing critical line. From left to right, we show the 3$\sigma$ Ly$\alpha$ emission from VLT/MUSE, the \jwst/NIRSpec IFU map combining \oiii\,$\lambda$5007 and \ha, and \jwst/NIRCam imaging in the short- and long-wavelength bands. The white ellipse in the MUSE and NIRSpec panels denotes the spectral extraction region for nebular emission analysis, while the dashed circles mark the eastern and western (SALSA\_E, SALSA\_W) compact emission components identified in the NIRSpec data. The NIRCam panels show the continuum emission in stacked images using the short- and long-wavelength bands, and the photometry is extracted within the corresponding cyan regions. The green curve indicates the critical line from the lens model at $z=5.66$.}
\label{fig:lens}
\end{figure*}

We also use archival \jwst/NIRCam images from programs GLASS-JWST Early Release Science \citet{treu_glass-jwst_2022} and UNCOVER \citet{bezanson_jwst_2024}, using v.7.2 of \texttt{grizli}-processed data products available through the DAWN JWST Archive (DJA). The observations are acquired using the filters F090W, F115W, F150W, F200W, in the short wavelength channel (SW) and F277W, F356W, F410M, and F444W in the long wavelength channel (LW) to perform photometry (see \S\ref{photometry}). Lastly, this cluster was previously observed with VLT/MUSE on 2014 September 21 and 2015 November 09 with a total integration time of 20 hours to detect Ly-$\alpha$ emission. A complete explanation on the MUSE data reduction is shown in \citet{richard_atlas_2021}. The imaging and spectroscopic observations described above are shown in Figure \ref{fig:lens}.

\subsection{NIRSpec Data Reduction}
\begin{figure*}[ht!]
\centering
\includegraphics[width=\linewidth]{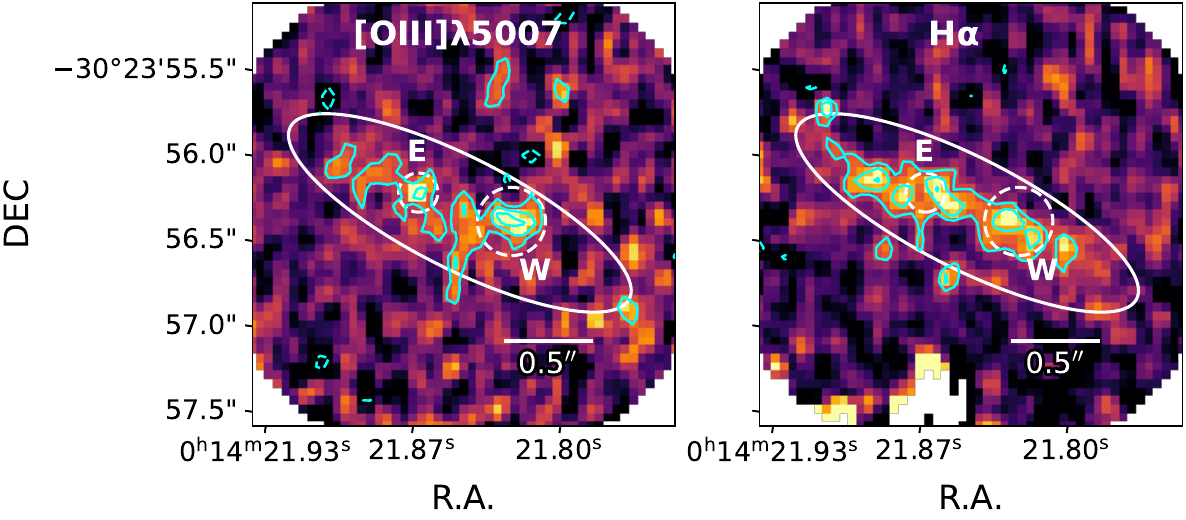}

\caption{Maps of the \oiii\,$\lambda$5007 (left) and H$\alpha$ (right) surface brightness. The maps are constructed after subtracting a local baseline estimated from adjacent line-free channels. Pixels with unreliable noise estimates are masked prior to plotting. The white ellipse indicates the aperture used to extract the one-dimensional spectrum, while the dashed circles SALSA\_E and SALSA\_W represent the clumps with highest signal-to-noise ratio. Cyan contours are shown at (-3, 3, 5, 7)$\sigma$ levels.}
\label{fig:maps}
\end{figure*}

Given that the targets in our NIRSpec IFU observations are some of the faintest objects ever studied, we performed two independent reductions to ensure robust results. Both methods yield consistent spectroscopy; one is described in Messa et al. (2026, in prep.), while the other is detailed here.
We use the \texttt{RegalJumper} data-reduction framework \citep{shajib_tdcosmo_2026}, which allows for flexible customization of different cleaning steps from the \jwst\ data reduction pipeline. For example, we iteratively refine the snowball rejection process to improve cosmic ray cleaning. We also investigate different replacement algorithms for flagged pixels. Instead of interpolating from neighboring pixels independently, we use the `adaptive\_trace\_model' option, which is better suited for our faint source as it optimally preserves the spectral line shape and flux. Artifact resampling is performed by replacing flagged pixels along the direction of the spectral trace in the 2D spectrum, with the resulting changes propagating into the noise and extraction. This algorithm option is implemented in Build 12.3 of the \jwst\ pipeline\footnote{\url{https://www.stsci.edu/contents/news/jwst/2026/pipeline-news-improved-correction-for-ifu-resampling-artifacts-in-build-12-3}}. The cube is constructed at a spatial sampling of 0\farcs05\,pixel$^{-1}$. Residual background is removed by selecting random empty regions of the sky.
The resulting surface brightness maps of the [\ion{O}{3}] and H$\alpha$ emission lines are shown in Figure~\ref{fig:maps}, where we identify the clumps SALSA\_E and SALSA\_W highlighted in dashed circles and detected at $\sim5-7\sigma$ significance.

\section{Lens Model}\label{lens}

The strong lensing model of the galaxy cluster Abell 2744 ($z=0.3072$) used in this work was developed with the \texttt{Gravity.jl} software \citep{lombardi_gravityjl_2024} building on the parametric model presented by \citet{bergamini_glass-jwst_2023}, obtained with \texttt{Lenstool} \citep{kneib_hubble_1996,jullo_bayesian_2007}. We refer the reader to \citet{bergamini_glass-jwst_2023} for a detailed description of the model and provide here only a brief summary. 

The observational constraints on which the model is based combine deep Hubble Space Telescope (\hst) imaging from the Hubble Frontier Fields \citep{lotz_frontier_2017} and Beyond Ultra-deep Frontier Fields And Legacy Observations \citep[BUFFALO;][]{steinhardt_buffalo_2020}, programs with \jwst/NIRCam observations from GLASS-JWST ERS, UNCOVER, and DDT program 2756 (PI: Chen). VLT/MUSE spectroscopy, in combination with the photometric data, was used by \citet{bergamini_glass-jwst_2023,bergamini_new_2023} to identify multiple-image systems and cluster member galaxies and to obtain their spectroscopic redshifts. 

Following \citet{bergamini_new_2023}, the projected total mass distribution of the cluster is modeled using four cluster-scale dark-matter halos, described with non truncated pseudo-isothermal elliptical mass distributions \citep[dPIE,][]{limousin_constraining_2005, eliasdottir07,bergamini_enhanced_2019}: two associated with the main cluster core and two accounting for external mass clumps in the surrounding cluster region. In addition, 177 cluster-member galaxies are included in the model. Of these, 172 are described as coreless spherical dPIEs, whose free parameters (i.e., the central velocity dispersion and truncation radius) follow luminosity-based scaling relations \citep[see Equation~5 in][]{bergamini_new_2023}. In total, the model includes 50 free parameters. 

The model is constrained by the observed positions of 149 multiple images, 121 of which are spectroscopically confirmed, associated with 50 background sources spanning the redshift range ($1.0258 \leq z \leq 9.756$). The model therefore has a total of 148 degrees of freedom. 

The best-fit model yields a root-mean-square separation of $\Delta_{\rm rms}=0\farcs36$ between the observed and model-predicted multiple-image positions, indicating a high precision in reproducing the strong lensing constraints.

At the position of SALSA, the predicted critical line from the model crosses through the emission as seen in Figure \ref{fig:lens}. The magnification is formally very high, making SALSA a very intrinsically faint system. Given the location of the critical line, the effective magnification depends on the exact surface brightness distribution in each band and line. We therefore estimated the magnification of the two clumps highlighted using 500 random realizations of the lens model. SALSA\_E lies closer to the critical line and consequently has a larger magnification and uncertainty yielding $\mu=151^{+114}_{-45}$, while SALSA\_W returns $\mu=69^{+10}_{-12}$, reflecting the strong sensitivity of the magnification to variations in the lens model in this region. Given the high magnification of the Eastern component and the position of SALSA relative to the critical line, we adopt $\mu>100$ as a conservative lower limit in our analysis.

\begin{figure*}[t]
    \centering
    \begin{subfigure}{0.34\textwidth}
        \includegraphics[width=\linewidth]{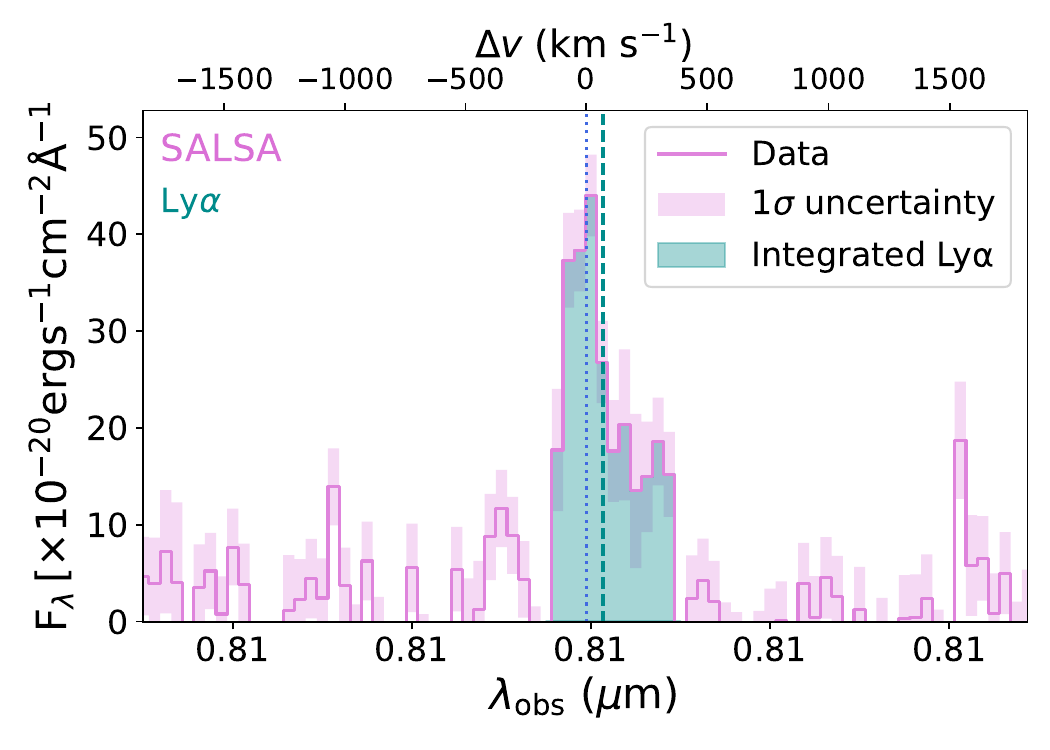}      
    \end{subfigure}
    \hfill
    \begin{subfigure}{0.32\textwidth}
        \includegraphics[width=\linewidth]{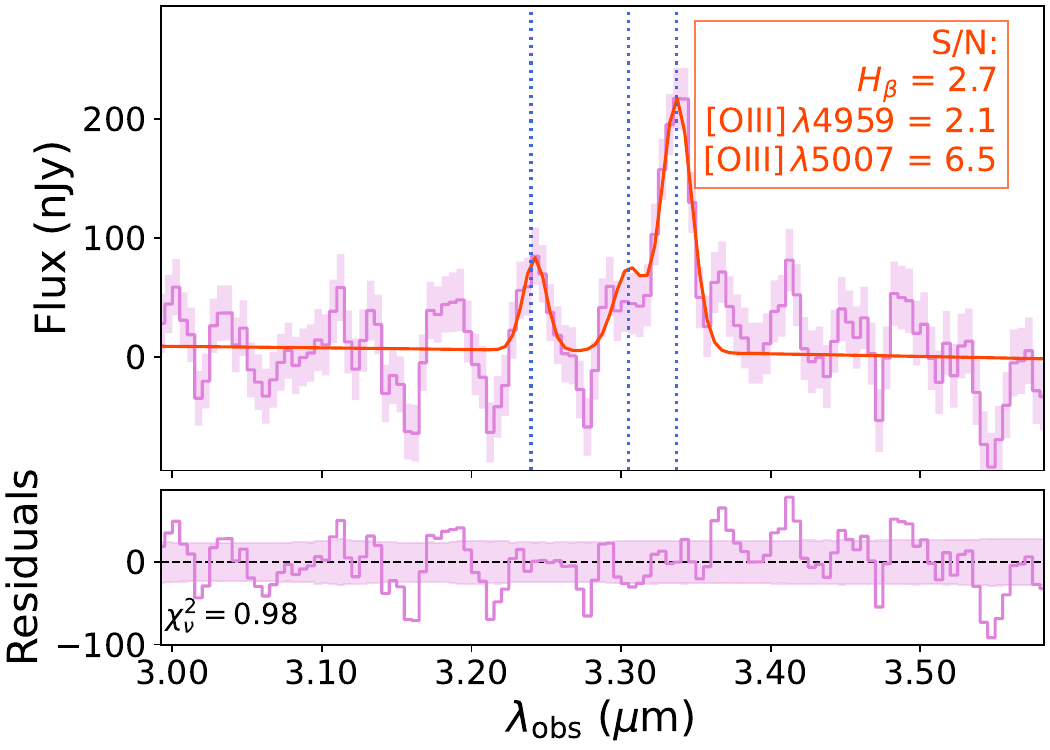}
    \end{subfigure}
    \hfill
    \begin{subfigure}{0.32\textwidth}
        \includegraphics[width=\linewidth]{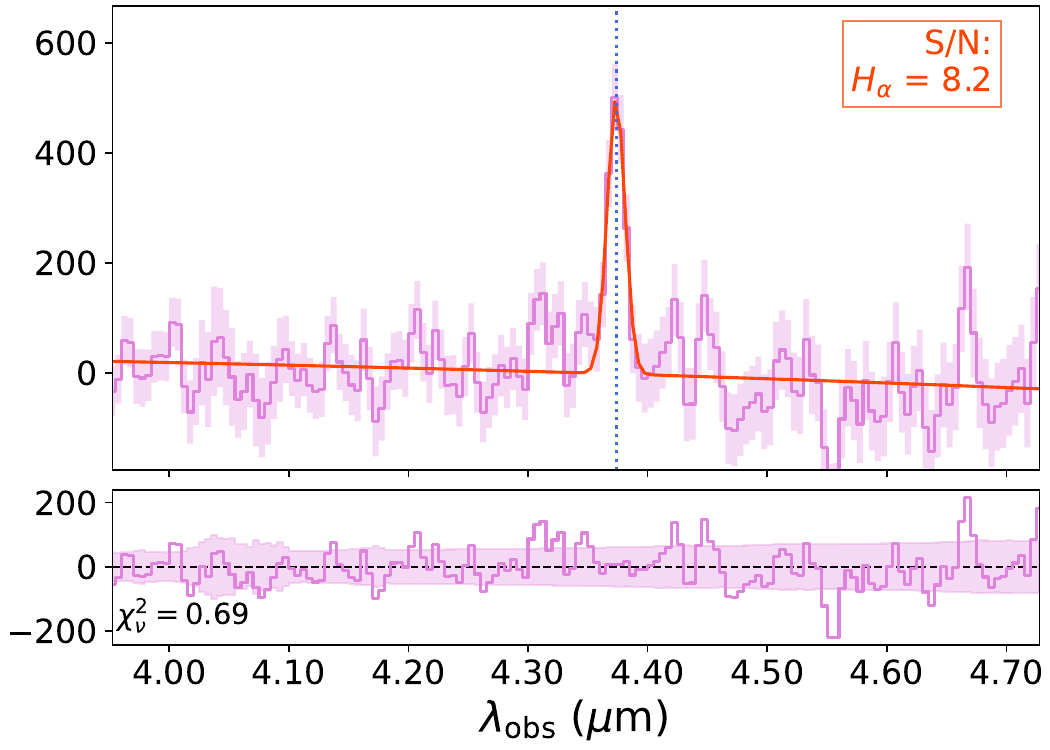}
    \end{subfigure}

    \vspace{0.1cm}
    
    \hspace*{\fill}
    \begin{subfigure}{0.32\textwidth}
        \includegraphics[width=\linewidth]{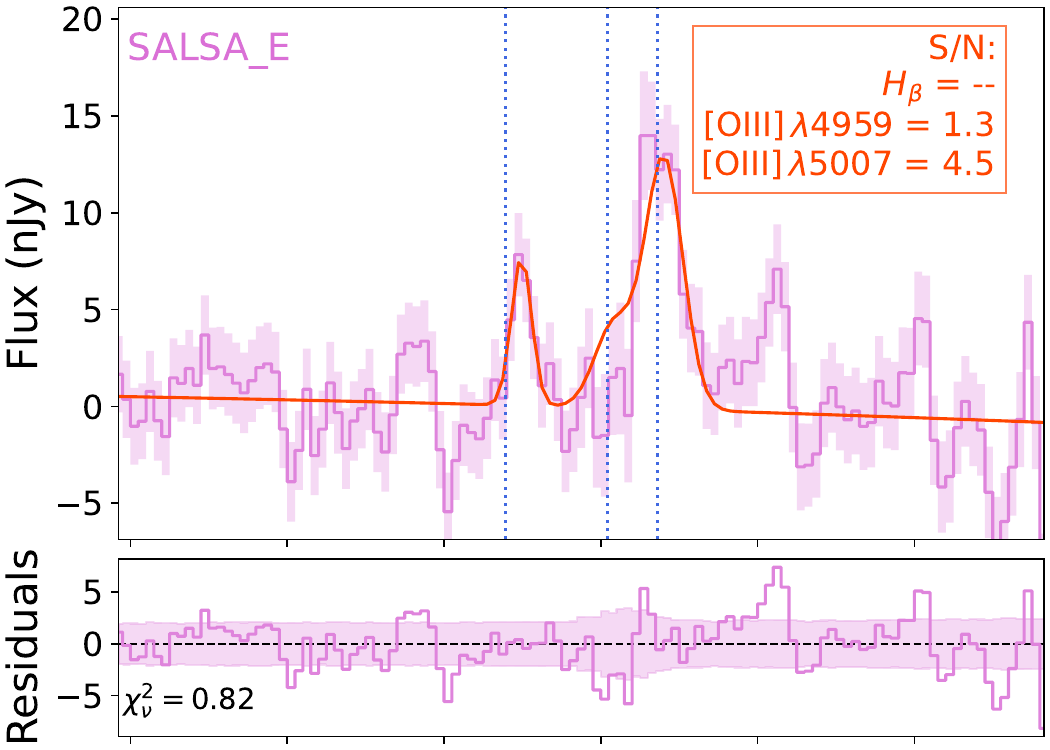}  
    \end{subfigure}
    \hspace{0.01\textwidth}
    \begin{subfigure}{0.32\textwidth}
        \includegraphics[width=\linewidth]{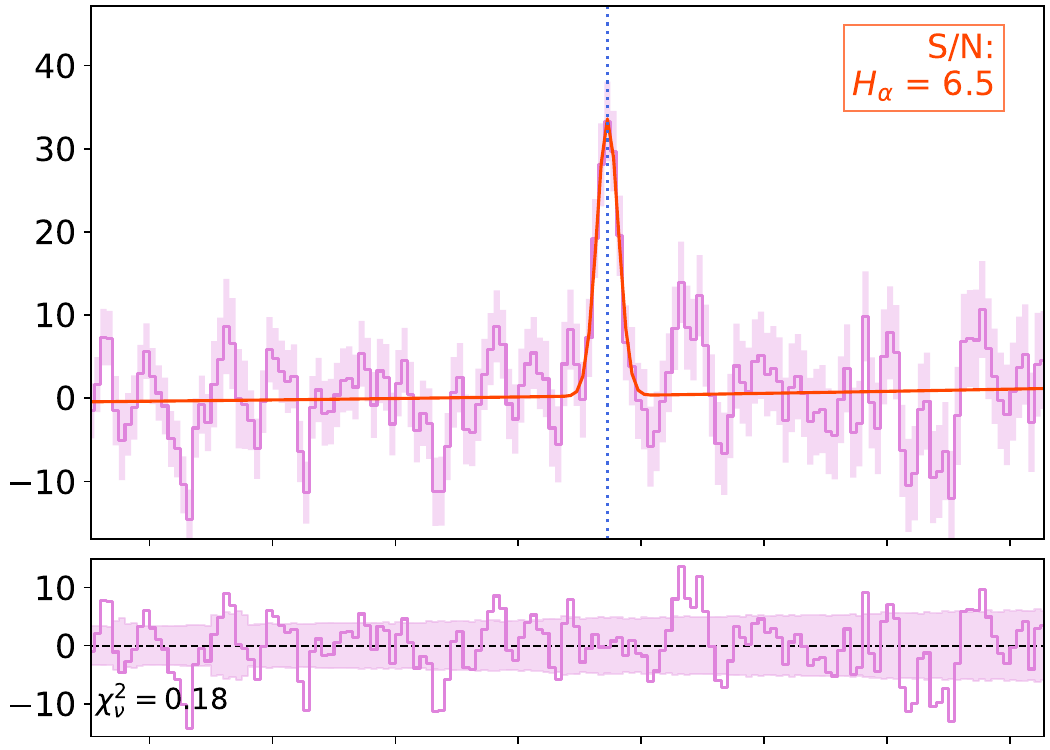}  
    \end{subfigure}
    \hspace*{\fill}

    \vspace{0.1cm} 
    \hspace*{\fill}
    \begin{subfigure}{0.32\textwidth}
    \centering
        \includegraphics[width=\linewidth]{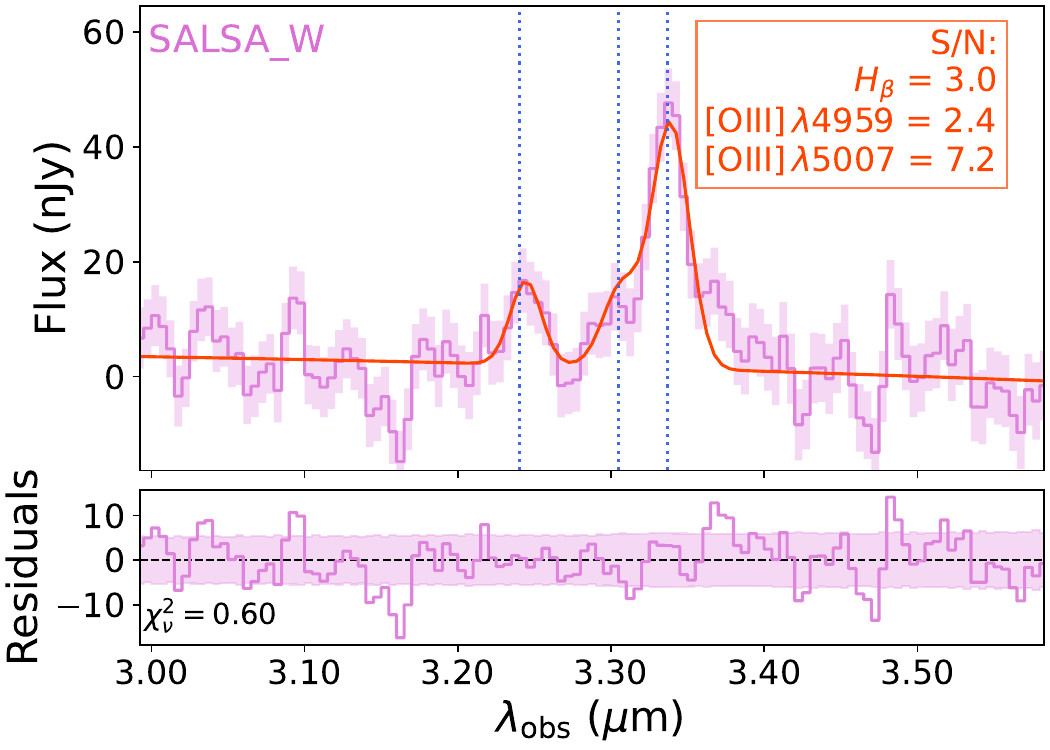}
    \end{subfigure}
    \hspace{0.01\textwidth}
    \begin{subfigure}{0.32\textwidth}
    \centering
        \includegraphics[width=\linewidth]{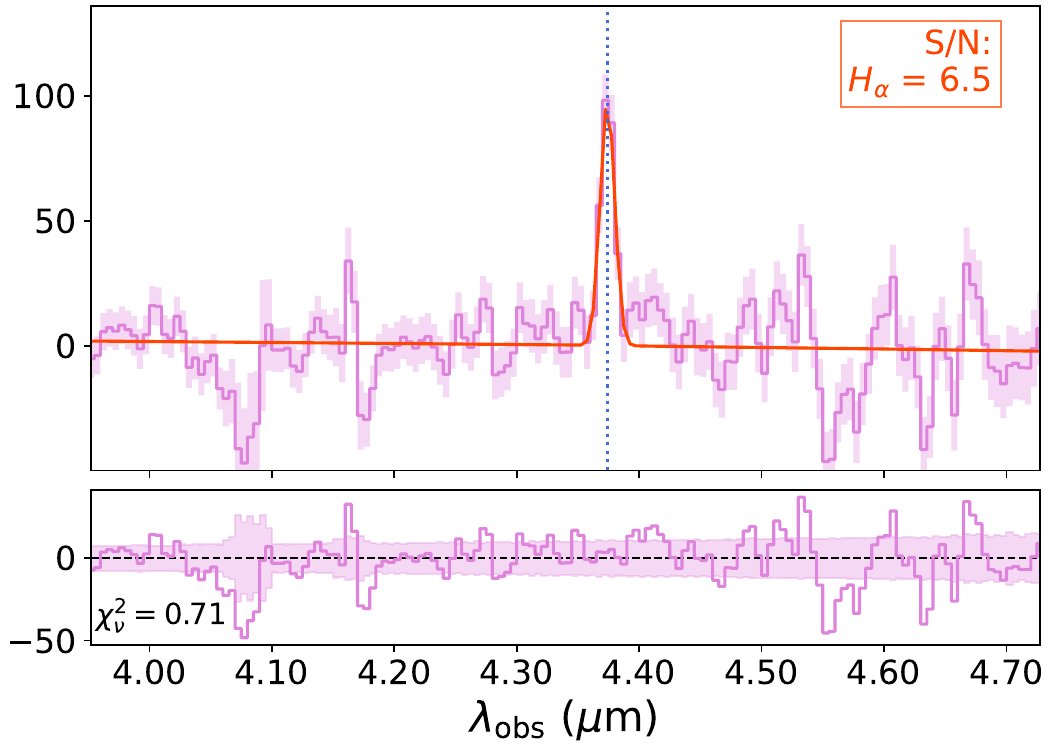}
    \end{subfigure}
    \hspace*{\fill}

    \caption{Spectral extraction from SALSA (\textit{top}) and the identified clumps SALSA\_E (\textit{middle}) and SALSA\_W (\textit{bottom}). The purple lines and shaded regions indicate the observed flux and 1$\sigma$ uncertainties. The Ly$\alpha$ spectrum of SALSA from MUSE is extracted from the same region as the H$\alpha$ emission in NIRSpec, with the integrated flux shown in shaded cyan region. The red lines show the Gaussian fits describing the emission lines centered at the blue dotted lines, as explained in \S\ref{emission-fit}. Signal-to-noise ratios for the emission lines correspond to their empirical local detection significance and are quoted left to right as they appear in the spectrum.}
    \label{fig:fits}
\end{figure*}

\section{Analysis}\label{analysis}
\subsection{Emission Line Fitting}\label{emission-fit}
We perform Gaussian fits for the emission lines of interest using two spectral windows, one covering H$\beta$ and \oiii\,$\lambda \lambda$4959,5007, and another for the H$\alpha$ emission. The latter, which has the highest signal-to-noise ratio (S/N) in the spectrum, is used to determine the systemic redshift of SALSA that yields $z=5.6626\pm0.0002$ and is then fixed when fitting the other emission lines. 

For each spectral window, the spectrum is modeled as a linear continuum together with Gaussian profiles using the \texttt{astropy.modeling} package. The emission line fits and derived quantities are calculated in 100 bootstrap realizations where the best-fit model is perturbed with Gaussian noise drawn from the pipeline pixel-to-pixel uncertainties. We noticed that the formal noise underestimates the observed scatter in the spectra, consistent with the discrepancy previously reported in previous NIRSpec observations \citep[e.g.,][]{ubler_ga-nifs_2023,fujimoto_alpine-cristal-jwst_2025}. Thus, for each realization, it is scaled by an empirical noise factor from the ratio of the standard deviation of the fit residuals to the median pipeline uncertainty within each window. The corresponding $1\sigma$ errors are calculated from the 16th and 84th percentiles of the bootstrapped distributions. The minimum S/N required for emission line detection is S/N$>3.0$. We note that only the emission lines \oiii\,$\lambda$5007 and H$\alpha$ comply with this criterion (see Figure~\ref{fig:fits}). Table~\ref{tabp} summarizes the best-fit parameters and physical properties of SALSA.

In the 2D emission-line maps shown in Figure \ref{fig:maps}, we identify two clumps highlighted in dashed circles detected at $\sim5-7\sigma$ significance in both \oiii\,$\lambda$5007 and \ha. We therefore extract the spectra from these regions to characterize whether they correspond to multiple images of the same source. We measure the \oiii\,$\lambda$5007 / H$\alpha$ ratios for the clump SALSA\_E $=0.95_{-0.14}^{+0.18}$, and SALSA\_W of $1.50_{-0.19}^{+0.26}$, which differ by $2\sigma$. This implies that we are observing different ionizing regions of the lensed galaxy at $z=5.66$, which agrees with the the lens model described in \S~\ref{lens} where both clumps are positioned within one side of the caustic and therefore distinct in nature (see Figure~\ref{fig:lens}).

\subsection{NIRCam Photometry}\label{photometry}
We perform photometry in sky-subtracted PSF-matched images to the NIRCam/F444W filter using an elliptical aperture enclosing the entire arc, see Figure~\ref{fig:lens}. As the arc has a very low surface brightness, we derive the rest-UV magnitude by stacking the observations in the SW bands F090W, F115W, F150W and F200W, obtaining $\rm mag_{SW}=28.0\pm0.2$. This value encloses within $\rm 1\sigma$ the values we would obtain by combining the photometry from individual filters. 
The intrinsic magnitude of SALSA is derived by considering the flux of the arc and correcting for lensing magnification, reaching $\rm M_{UV}>-13.6$. This limit places SALSA among the faintest galaxies ever observed at similar redshifts \citep{atek_glimpse_2026}. Similarly, we calculated the apparent and absolute UV magnitudes for the compact regions SALSA\_E with $\rm mag_{SW}=30.1\pm0.2$, $\rm M_{UV}>-11.5$ and SALSA\_W $\rm mag_{SW}=29.7\pm0.2$, $\rm M_{UV}>-11.9$.
Notably, the nebular emission detected in the NIRSpec IFU data for SALSA\_E has no clear stellar continuum counterpart (see Figure \ref{fig:lens}). This is consistent with the emission being powered by a very young stellar population for which the ionizing emission can be prominent before a significant continuum contribution becomes detectable. However, given the faintness of the continuum, this interpretation remains tentative.

We estimate SALSA's rest-optical magnitude by stacking the LW bands, finding $\rm mag_{LW}=29.1\pm0.4$, corresponding to $\rm M_{opt}>-12.6$. The continuum flux measured in the broadband filters is contaminated by the line flux emission (though the stacking of the bands gives a very broad baseline); we can take the measured magnitude as an upper limit on the continuum flux, and use the \ha\ flux to derive $\rm EW(H\alpha)>1600~\AA$, pointing towards a young age for this source $\rm <5~Myr$ in the single-burst assumption for standard stellar models \citep[e.g.,][]{leitherer_starburst99_1999,bruzual_stellar_2003,eldridge_binary_2017}. Assuming an age of $\rm \approx3~Myr$, the rest-optical continuum flux limit would convert\footnote{Assuming BPASS stellar models \citep{eldridge_binary_2017} with a \citet{kroupa_variation_2001} initial mass function, an instantaneous-burst star-formation history and a $\rm 5\%~Z_\odot$ metallicity.} into a limit on the stellar mass of the system $\rm M<10^5~M_\odot$.
The large uncertainties arising from the low signal to noise ratio of the continuum in the NIRCam bands, especially in the long wavelength channel, prevent from deriving more accurate mass estimates. 

The angular extension of the arc ($\sim1\farcs6$) can be used to derive an estimate of the intrinsic radius, $\rm R\lesssim100$pc measured assuming a tangential magnification $\mu_{tan}>50$; we caution the reader that the observed arc could be only a lensed portion of a larger galaxy. 

\subsection{MUSE Ly$\alpha$ Spectrum}\label{Lya}
We measure the Ly$\alpha$ emission from SALSA using the same aperture as H$\alpha$, ensuring a consistent spatial extraction. To estimate the total Ly$\alpha$ flux while accounting for seeing losses, an aperture correction is applied based on a normalized Gaussian profile with the measured seeing of the MUSE observations. The central wavelength is calculated from the flux-weighted centroid of the sky-subtracted emission (cyan shaded region in Figure \ref{fig:fits}), and is found to be marginally redshifted relative to the \ha\ systemic redshift $\Delta$v$_{Ly\alpha} = 14.5_{-9.5}^{+12.4}$\,km~s$^{-1}$, consistent with zero velocity offset within the uncertainties. This is among the smallest velocity offsets measured at these redshifts, comparable to the lowest values reported in \citet{messa_faint_2026} (see their Figure~A1 and Table~A.1). Such a small offset could point towards LyC leaking.

We calculate the flux ratio Ly$\alpha$/H$\alpha=3.42_{-1.23}^{+1.25}$, below the Case B recombination value of 8.7, corresponding to an apparent Ly$\alpha$ escape fraction of $f_{\rm{esc}}^{\mathrm{Ly}\alpha}=0.39\pm0.14$ under the assumption of negligible differential dust attenuation and that the adopted aperture captures the Ly$\alpha$ emission, which may be more spatially extended owing to resonant scattering. SALSA's Ly$\alpha$ escape fraction is high in comparison to typical values found at similar redshifts. Recent studies combining Ly$\alpha$ and H$\alpha$ measurements for galaxies find generally modest Ly$\alpha$ escape fractions at $z\sim5-6$ with population averages or medians of $\sim0.1$--$0.3$ \citep[e.g.,][]{chen_jwst_2024,lin_quantifying_2024,tang_ly_2024,shimizu_subaru_2026}. However, these studies find individual sources with substantially higher escape fractions reaching $f_{\rm{esc}}^{\mathrm{Ly}\alpha}>0.6$. Therefore, SALSA lies towards the high end of the distribution, but is not exceptional compared to the typical galaxy population at $z\gtrsim5$.


\begin{deluxetable}{lc}
\tablecaption{Physical Properties for SALSA\label{tabp}}
\tablehead{
\colhead{Quantity}
& \colhead{SALSA} 
}
\startdata
RA & 0:14:21.8678\\
DEC & -30:23:56.241\\
Redshift\,(H$\alpha$) & $5.6626\pm 0.0002$ \\
$\mathrm{Ly}\alpha\ [10^{-20} \rm{erg~s}^{-1}\mathrm{cm}^{-2}$] & $502\pm180$\\
$\mathrm{H}\alpha\ [10^{-20} \rm{erg~s}^{-1}\mathrm{cm}^{-2}$] & $147.0_{-14.0}^{+10.4}$\\
$\mathrm{[O\,III]}\,\lambda5007$ [10$^{-20} \rm{erg~s}^{-1}\mathrm{cm}^{-2}$] & $146.7_{-13.4}^{+10.4}$\\
$\mathrm{R3} = \frac{\mathrm{[O\,III]}\,\lambda5007}{H\alpha / 2.86}$\tablenotemark{a} & $2.82_{-0.25}^{+0.34}$ \\
$\mathrm{12+log(O/H)}$\tablenotemark{b} & $7.43\pm0.09$ \\
$\mu_{\rm{tot}}~(\mu_{\rm{tan}})$\tablenotemark{c} & $>100~(>50)$ \\
M$_{\rm{UV}}$\tablenotemark{d} & $ > -13.6$ \\
L$_{\rm{UV}}$ [$10^{25} \rm{erg~s}^{-1}\rm{Hz}^{-1}$]\tablenotemark{d} & $< 2.50$ \\
L$_{\rm{H\alpha}}$ [$10^{39} \rm{erg~s}^{-1}$]\tablenotemark{d} & $5.12_{-0.49}^{+0.36}$ \\
$\mathrm{SFR}(\mathrm{UV})\ [10^{-2} \rm{M}_\odot\,\mathrm{yr^{-1}}]$\tablenotemark{d} & $1.71_{-0.29}^{+0.35}$ \\
$\mathrm{SFR}(\mathrm{H}\alpha)\ [10^{-2} \rm{M}_\odot\,\mathrm{yr^{-1}}]$\tablenotemark{d} & $4.05_{-0.39}^{+0.29}$ \\
log($\xi_{\rm{ion}})$[Hz\,erg$^{-1}$] & $25.49_{-0.08}^{+0.09}$ \\
M$_{\rm{UV}}\_E,\_W$  & $ >-11.5~,~>-11.9$ \\
log($\xi_{\rm{ion}}\_E,\_W)$[Hz\,erg$^{-1}$] & $25.24_{-0.09}^{+0.10}$~,~$25.38_{-0.1}^{+0.1}$ \\
$\mathrm{R3\_E,\_W}$ & $2.63_{-0.23}^{+0.40}$~,~$4.43_{-0.27}^{+0.73}$ \\
$\mathrm{12+log(O/H)\_E,\_W}$ & $7.39_{-0.09}^{+0.10}$~,~$7.67\pm0.11$ \\
\enddata
\tablenotetext{a}{Computed assuming Case B recombination ratio H$\alpha$/H$\beta=$2.86 with no dust attenuation.}
\tablenotetext{b}{Calculated using the R3 relation in \cite{curti_mass-metallicity_2020} and assuming $Z_\odot = 8.69$ \citep{asplund_chemical_2009}.}
\tablenotetext{c}{Lower limit on the total and tangential magnification factor calculated in this study (see \S\ref{lens}).} 
\tablenotetext{d}{Values calculated after correcting for the total magnification factor $\mu_{\rm{tot}}$.}

\end{deluxetable}

\section{Ionizing Photon Production Efficiency and Metallicity}\label{ionizing}
We derive the ionizing photon production efficiency $\xi_{\rm ion}$ from SALSA and the brightest regions SALSA\_E and SALSA\_W to investigate the ionizing properties of this faint system and compare them to the literature at similar cosmic epochs. The total ionizing photon production rate, Q(H$^0$), is inferred from the H$\alpha$ luminosity following \citet{leitherer_synthetic_1995} which assumes Case B recombination in an ionization-bounded nebula with $\rm{T_e}\sim 10^4$K, a Salpeter IMF \citep{salpeter_luminosity_1955}, and negligible escape of LyC ionizing photons ($f_{\rm{esc}}^{\rm{LyC}}=0$), see Equation \ref{eq1}. Under these assumptions, the \ha\ luminosity traces the rate of hydrogen-ionizing photon production.

\begin{equation}
    \xi_{\rm ion} = \frac{Q(\rm{H}^0)}{L_{UV}}
    = \frac{L(\rm{H}_\alpha) / 1.36\times10^{-12}}{L_{UV}} \quad [\rm{Hz}~\rm{erg}^{-1}]
    \label{eq1}
\end{equation}

For SALSA, we find a high ionizing photon production efficiency of log($\xi_{\rm{ion}}$) $= 25.49_{-0.36}^{+0.52}$ Hz\,erg$^{-1}$. Similarly, for the clumps SALSA\_E and SALSA\_W, we obtain log($\xi_{\rm{ion}}\_E$)$= 25.24_{-0.09}^{+0.10}$ Hz\,erg$^{-1}$ and log($\xi_{\rm{ion}}\_W) = 25.38_{-0.1}^{+0.1}$ Hz\,erg$^{-1}$, respectively. These values are broadly consistent with the commonly adopted canonical range of log($\xi_{\rm{ion}}$)$ = 25.2-25.3$ predicted by stellar population models \citep[e.g.,][]{robertson_cosmic_2015,robertson_galaxy_2022}, although SALSA itself lies slightly above this range. 

\begin{figure}
\centering
\includegraphics[width=\linewidth]{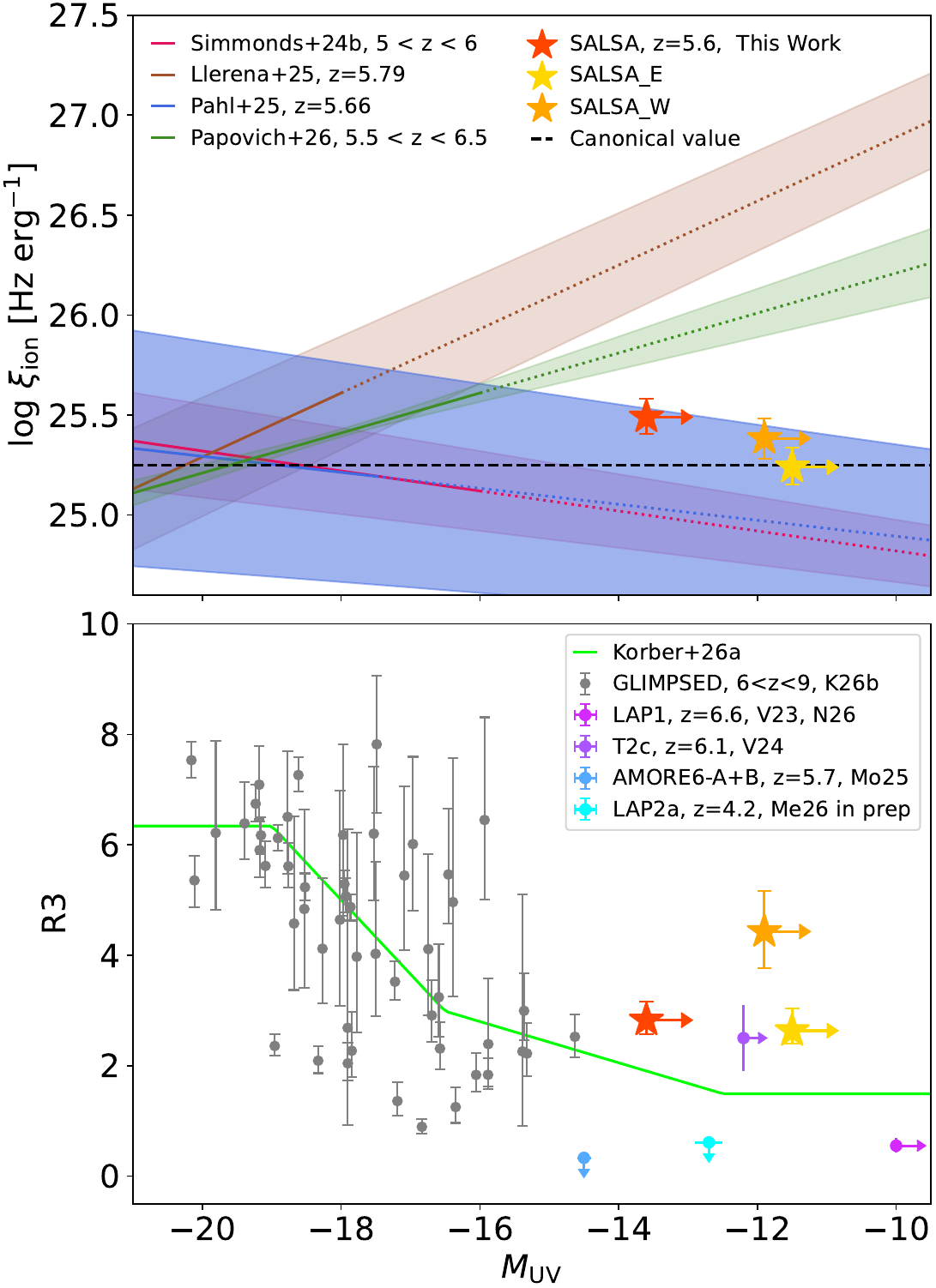}
\caption{$\xi_{\rm{ion}}$ and R3 ratio as a function of UV brightness for the sources presented in this work, shown as stars. \textit{Top:} Comparison to $\xi_{\rm{ion}}$ with M$_{\rm UV}$ relations from recent \jwst\ studies \citep{simmonds_low-mass_2024,llerena_ionizing_2025,pahl_spectroscopic_2025,papovich_galaxies_2026}, together with the canonical value of log($\xi_{\rm{ion}}$)$=25.2$ \citep{robertson_discovery_2022}. \textit{Bottom:} The green line shows the R3M$_{\rm UV}$ relation assumed in \citet{korber_glimpse_2026}. Literature measurements of comparatively low-luminosity sources at $z>4$ are shown for comparison, including GLIMPSED \citep[K26b:][]{korber_glimpse-d_2026}, LAP1 \citep[V23,N26:][]{vanzella_extremely_2023,nakajima_ultra-faint_2026}, T2c \citep[V24:][]{vanzella_extreme_2024}, AMORE6-A+B \citep[Mo25:][]{morishita_pristine_2025}, and LAP2a (Me26: Messa et al.~2026, in prep.)}
\label{fig:xi_ion}
\end{figure}

We compare our measurements to predicted values from extrapolations of empirical relations in the literature at similar redshifts but caution the reader that all these relations do not reach magnitudes fainter than M$_{\rm{UV}}\sim-16$ (see Figure \ref{fig:xi_ion}). SALSA and the clumps are within $1\sigma$ of the prediction by Equation~5 in \citet{pahl_spectroscopic_2025}, as well as the median value of log($\xi_{\rm{ion}}$)$=25.47_{-0.22}^{+0.61}$ Hz\,erg$^{-1}$ at $z>4$ in \citet{pahl_aurora_2026}. Extrapolating the relation by \citet{simmonds_ionizing_2024} results in an under-prediction by $1.60-1.85$~dex for our measurements.

In contrast, SALSA and the clumps SALSA\_E and SALSA\_W are $>3\sigma$ ($0.81$, $1.06$ and $0.92$ dex, respectively) below the prediction from the $z=5.79$ relation found in \citet{llerena_ionizing_2025}, although their sample is complete only to M$_{\rm{UV}}=-18$). Similarly, comparing to the $5.5<z\leq6.5$ relation by \citet{papovich_galaxies_2026} results in SALSA and the clumps lying $0.36-0.61$ dex below this prediction. These comparisons illustrate how sensitively the inferred $\xi_{\rm{ion}}$--UV luminosity relation depends on sample selection and dust correction assumptions to the nebular emission on the Balmer decrement. These may contribute substantially to the dispersion in inferred $\xi_{\rm{ion}}$, or flattening of the relation towards faint luminosities. The paucity of comparisons underscores the importance of obtaining additional measurements of $\xi_{\rm{ion}}$ in this low luminosity range.

The measured $\xi_{\rm{ion}}$ for SALSA and the clumps are broadly consistent with the range of theoretical stellar population models including very massive stars (VMS; $>100$\Msun) explored in  \citet{schaerer_observable_2025}. Our measurements lie between the equilibrium constant-SFR value of log($\xi_{\rm{ion}}$)$ = 25.34$ Hz\,erg$^{-1}$ for their VMS-400\Msun\ model and the maximum value of log($\xi_{\rm{ion}}$)$=25.80$ Hz\,erg$^{-1}$ reached by a young burst. In contrast, the extreme Pop III IMF C model predicts substantially higher values, log($\xi_{\rm{ion}}$)$=26.08$ Hz\,erg$^{-1}$ for constant SFR and 26.18 Hz\,erg$^{-1}$ for a burst. Thus, our measurement is consistent with young stellar populations containing VMS without requiring the extreme top-heavy IMF associated with the Pop III stars.

The high ionizing photon production of SALSA relative to the UV luminosity may arise from a very young or rapidly evolving stellar population. To investigate this, we compare the SFR inferred from \ha\ and the rest-frame UV, which probe different timescales of recent star formation. The SFR(\ha) $=4.00_{-0.39}^{+0.29} \times 10^{-2}\rm{M}_\odot\,\mathrm{yr^{-1}}$ and SFR(UV)$ = 1.71_{-0.29}^{+0.35} \times 10^{-2}\rm{M}_\odot\,\mathrm{yr^{-1}}$ using the calibration in \citet{kennicutt_star_1998} and a Salpeter IMF over 0.1-100\Msun. Because $\mu>100$, these values represent the maximum intrinsic SFRs allowed by the adopted magnification. The resulting ratio SFR(\ha)/SFR(UV)$\sim2.3$ suggests that the current ionizing-star formation inferred from \ha\ is elevated relative to the longer-timescale star formation traced by the UV continuum, consistent with a recent burst or rapidly rising star-formation history.

To characterize the ionization conditions and nebular excitation of SALSA, we investigate the relation between the line ratio R3 (\oiii\,$\lambda$5007/H$\beta$) and UV luminosity. Given that the \hb\ emission is not significantly detected in our observations (see \S \ref{emission-fit}), we use H$\alpha$ and assume Case B recombination with the intrinsic ratio H$\beta$/H$\alpha =2.86$ for zero dust attenuation (A$_V = 0$) \cite{storey_recombination_1995, osterbrock_astrophysics_2006}, which yields R3$ = 2.82_{-0.25}^{+0.34}$. Currently, the study reaching the faintest galaxies and investigating their relation to R3 is GLIMPSE-D \citep{korber_glimpse_2026,korber_glimpse-d_2026}. Relative to this relation, our R3 measurement is $2.7\sigma$ higher than the expected value. Together with measurements of other faint galaxies (see Figure \ref{fig:xi_ion}) this suggests that R3 may exhibit larger scatter at the lowest UV-luminosities (i.e. not all low-luminosity galaxies have extremely low R3). The excess R3 value for SALSA may indicate enhanced nebular excitation relative to the low-luminosity population; however, this diagnostic is degenerate with both metallicity and the ionizing radiation field. The high $\xi_{\rm{ion}}$ of SALSA also provides independent evidence for the efficient production of ionizing photons, which may contribute to its high R3 measurement.

The R3 ratio is also sensitive to the gas-phase oxygen abundance, with higher R3 values generally associated with lower metallicities, although the precise relation also depends on the ionization conditions \citep[e.g.,][]{laseter_jades_2024,korber_glimpse-d_2026}. We therefore derive an R3-based oxygen abundance estimate following the calibration of \citet{curti_mass-metallicity_2020} and find $12+\rm{log(O/H)} = 7.43\pm0.09$, or $\sim5.5$\% of the solar value. Notably, the clump SALSA\_W has a higher R3 ratio $4.43_{-0.27}^{+0.73}$ driven by its comparatively stronger \oiii\,$\lambda$5007 emission (see Figure~\ref{fig:maps}), placing it $3.4\sigma$ above the \citet{korber_glimpse_2026} prediction. Its oxygen abundance is $12+\rm{log(O/H)} = 7.67\pm0.11$, corresponding to $\sim9.5$\% of the solar value.

\section{Summary}\label{summary}
This Letter presents SALSA, an intrinsically faint source at $z=5.66$ with a high ionizing power log($\xi_{\rm{ion}}$)$=25.49_{-0.08}^{+0.09} $~Hz\,erg$^{-1}$, low metallicity $\mathrm{12+log(O/H)}= 7.43\pm0.09$, high R3 index $2.82_{-0.25}^{+0.34}$, and high Ly$\alpha$ escape fraction $f_{\rm{esc}}^{\mathrm{Ly}\alpha}=0.39\pm0.14$. The NIRSpec IFU observations demonstrate the sensitivity of \jwst\ to shed light on extremely faint sources and investigate their contribution to Reionization. The combination of high $\xi_{\rm{ion}}$, elevated R3 and low metallicity provides complementary evidence for an intense ionizing radiation field and highly excited nebular and metal-poor ionized gas. These findings support a scenario in which the abundant population of faint galaxies can efficiently produce the majority of ionizing photons responsible for Hydrogen Reionization. SALSA is one of the multiple arclets found through lensing. Future observations of larger samples of such extreme systems, particularly with higher spectral and spatial resolution, will be essential for determining how representative their properties are and placing faint galaxies in the broader context of galaxy evolution and Reionization.

\begin{acknowledgments}
This work is based on observations made with the NASA/ESA/CSA James Webb Space Telescope. The NIRSpec/IFU observations under program \# 7677 were obtained from the Mikulski Archive for Space Telescopes (MAST) and can be accessed via \dataset[doi: 10.17909/dst0-sp22]{https://doi.org/10.17909/dst0-sp22}. Support for SRR and TT in program JWST-GO-7677 was provided by NASA through a grant from the Space Telescope Science Institute, which is operated by the Association of Universities for Research in Astronomy, Inc., under NASA contract NAS5-03127. EV and MM acknowledge financial support through INAF GO Grant 2024 ``Mapping Star Cluster Feedback in a Galaxy 450 Myr after the Big Bang'' and the project PRORIS - COSMOWEB ``A new era for cosmology:  exploiting the JWST revolution''. PB acknowledges financial support from the Italian Space Agency (ASI) through contract ``Euclid - Phase E'', INAF Grants ``The Big-Data era of cluster lensing'' and ``Probing Dark Matter and Galaxy Formation in Galaxy Clusters through Strong Gravitational Lensing''. ML acknowledges support from the National
Recovery and Resilience Plan (NRRP), funded by the European Union – NextGenerationEU; Project title ``GRAVITY”, project code PNRR\_BAC24MLOMB\_01, CUP C53C22000350006.
\end{acknowledgments}

\facilities{\jwst, VLT}

\software{Astropy \citep{the_astropy_collaboration_astropy_2013}, Gravity.jl \citep{lombardi_gravityjl_2024}, Matplotlib \citep{hunter_matplotlib_2007}, Numpy \citep{harris_array_2020}, RegalJumper \citep{shajib_tdcosmo_2026}, SciPy \citep{virtanen_scipy_2020}
          }

\bibliography{references,other_referenced}{}
\bibliographystyle{aasjournalv7}

\end{document}